\documentclass[pdflatex,sn-mathphys-num]{sn-jnl}

\usepackage{graphicx}%
\usepackage{multirow}%
\usepackage{amsmath,amssymb,amsfonts}%
\usepackage{amsthm}%
\usepackage{mathrsfs}%
\usepackage[title]{appendix}%
\usepackage{xcolor}%
\usepackage{textcomp}%
\usepackage{manyfoot}%
\usepackage{booktabs}%
\usepackage{algorithm}%
\usepackage{algorithmicx}%
\usepackage{algpseudocode}%
\usepackage{listings}%
\usepackage{lineno}
\usepackage{subcaption}
\usepackage{booktabs}
\usepackage{tabularx}
\usepackage{acro}
\usepackage[capitalise,noabbrev]{cleveref}
\usepackage{placeins}

\newcommand\rev[1]{\textcolor{black}{#1}}
\newcommand\revTwo[1]{\textcolor{black}{#1}}
\newcommand\revThree[1]{\textcolor{black}{#1}}

\DeclareAcronym{iou}{
  short=IoU,
  long=intersection over union,
}
\DeclareAcronym{cas}{
  short=CAS,
  long=computer-aided surgery,
}
\DeclareAcronym{tka}{
  short=TKA,
  long=Total Knee Arthoplasty,
}

\begin{document}

\title[Sound Source Localization for Spatial Mapping of Surgical Actions in Dynamic Scenes]{Sound Source Localization for Spatial Mapping of Surgical Actions in Dynamic Scenes}

\author[1,3]{\fnm{Jonas} \sur{Hein}}
\author[2]{\fnm{Lazaros} \sur{Vlachopoulos}}
\author[2]{\fnm{Maurits Geert Laurent} \sur{Olthof}}
\author[1]{\fnm{Bastian} \sur{Sigrist}}
\author[1]{\fnm{Philipp} \sur{Fürnstahl}}
\author*[1]{\fnm{Matthias} \sur{Seibold}}\email{matthias.seibold@balgrist.ch}

\affil[1]{\orgdiv{ROCS}, \orgname{Balgrist University Hospital, University of Zurich}, \country{Switzerland}}
\affil[2]{\orgdiv{Dept. of Orthopedics}, \orgname{Balgrist University Hospital, University of Zurich}, \country{Switzerland}}
\affil[3]{\orgdiv{Computer Vision and Geometry Group}, \orgname{ETH Zurich}, \country{Switzerland}}

\abstract{\textbf{Purpose:} Surgical scene understanding is key to advancing computer-aided and intelligent surgical systems. Current approaches predominantly rely on visual data or end-to-end learning, which limits fine-grained contextual modeling. This work aims to enhance surgical scene representations by integrating 3D acoustic information, enabling temporally and spatially aware multimodal understanding of surgical environments.

\textbf{Methods:} We propose a novel framework for generating 4D audio-visual representations of surgical scenes by projecting acoustic localization information from a phased microphone array onto dynamic point clouds from a RGB-D camera. A transformer-based acoustic event detection module identifies relevant temporal segments containing tool–tissue interactions which are spatially localized in the audio-visual scene representation. The system was experimentally evaluated in a realistic operating room setup during simulated surgical procedures performed by experts.

\textbf{Results:} The proposed method successfully localizes surgical acoustic events in 3D space and associates them with visual scene elements. Experimental evaluation demonstrates accurate spatial sound localization and robust fusion of multimodal data, providing a comprehensive, dynamic representation of surgical activity.

\textbf{Conclusion:} This work introduces the first approach for spatial sound localization in dynamic surgical scenes, marking a significant advancement towards multimodal surgical scene representations. By integrating acoustic and visual data, the proposed framework enables richer contextual understanding and provides a foundation for future intelligent and autonomous surgical systems.}


\maketitle

\section{Introduction}
Surgical procedures are inherently dynamic, involving complex interactions among surgical staff, patients, instruments, and other medical devices. 
\rev{To enable the next generation of intelligent \ac{cas} systems, capable of context-aware assistance or semi-autonomous operation, machines must possess a holistic understanding of this environment. 
This capability, known as surgical scene understanding, relies on creating digital representations that accurately mirror the physical reality of the operating theater.}

\rev{Historically, the field has prioritized visual perception. 
A body of works proposed to explicitly model the surgical context based on RGB or RGB-D data, employing semantic structures such as action triplets \cite{nwoye2020triplets} or scene graphs \cite{oeszoy2022ssg} to capture relationships between surgeons, tools, and tissues.}
In contrast to explicit modeling approaches, end-to-end training has recently demonstrated strong performance in predicting high-level surgical concepts (e.g., surgical phases) from low-level observations (e.g., individual video frames) \cite{Ozsoy_2025_CVPR,oezsoy2025egoexoregoexocentricoperatingroom}. 
\rev{Nevertheless, these visual-centric approaches face inherent limitations. 
Visual data is susceptible to occlusion, lighting variations, and often fails to capture the physical nature of tool-tissue interactions, such as the mechanical resistance of bone during sawing or the precise moment of a drill breakthrough.}

\rev{Multimodal surgical perception has emerged as a promising approach to overcome the limitations of unimodal vision. 
Several works proposed digital twins of surgery, which integrate complementary data streams such as kinematics, inertial measurements and system events into a comprehensive digital representation of the surgical scene, effectively decoupling low-level sensing from high-level semantic reasoning \cite{ding2024dt, Hein_2024_CVPR}. 
Within this multimodal landscape, audio represents a largely untapped but highly information-rich modality. 
}

\rev{Surgical sounds provide high-frequency cues about physical interactions that are often invisible to cameras.}
Recently, acoustic events generated during tool–tissue interactions—such as sawing, chiseling, or drilling—have been shown to provide valuable insights into surgical tasks \cite{seibold2022conditional, goossens2020acoustic, seibold2021realtime}. 
Furthermore, in our own previous work, we proposed a proof-of-concept for the localization of acoustic events in surgery based on a microphone array and acoustic beamforming in 2D image space \cite{seibold2024ssl}. 
\rev{The integration of acoustic information into digital surgical scene representations introduces unique capabilities, such as high temporal resolution (sub-fps) and the ability to detect events that cannot be captured by vision alone (e.g., activity of power tools), but requires spatial and temporal integration with the visual representation to enable effective multimodal surgical scene understanding.
To the best of the author’s knowledge, there is currently no comparable approach that integrates spatial sound into dynamic surgical scene representations.}

In this paper, we propose a novel paradigm to enable sound localization in dynamic surgical scenes, thereby enhancing digital surgical scene representations to achieve improved contextual understanding for intelligent surgical systems. 
The key contributions of this work can be summarized as follows:

\begin{itemize}
    \item \rev{We propose a novel concept to generate 4D audio-visual representations of surgical scenes defined by a time-varying visual geometric 3D representation of a surgical scene that is enriched with localized acoustic events occurring at certain points of time or spanning over a defined period. Our approach achieves this by fusing dynamic point clouds obtained using a RGB-D camera with acoustic localization information obtained from \revTwo{a} phased microphone array, also known as acoustic camera.}
    \item To identify relevant frames for event localization, we propose an acoustic event detection stage that is based on a transformer architecture and enables the detection of surgical acoustic events in a continuous surgical sequence to trigger the localization of detected acoustic events.
    \item We thoroughly evaluate the proposed method with experimental data that has been recorded from a simulated surgical procedure executed by surgical experts in a realistic surgical environment.
\end{itemize}

\revTwo{The code and associated data are available under \url{https://github.com/matthiasseibold/ssl_v2}.}

\section{Related Work}
\rev{The proposed approaches for surgical scene understanding can be categorized by the modalities used to construct the digital representation, namely purely visual approaches, acoustic sensing, and multimodal fusion. The majority of existing works relies on optical data such as RGB or RGB-D to derive both semantic and spatial information for modeling the surgical context. 
Early approaches focused on temporal and hierarchical  modeling, using surgical process models \cite{neumuth2017spm} to extract phases and steps from video data \cite{padoy2019workflow}. 
More recently, spatial understanding has been enhanced through action triplets (instrument, action, target) \cite{nwoye2020triplets} and surgical scene graphs, which model the operating room as a structured graph of actors and objects \cite{oeszoy2022ssg, oeszoy2023labrador, hamoud2024stor}.
While end-to-end visual models achieve strong performance on specific benchmarks \cite{Ozsoy_2025_CVPR}, they lack the ability to capture non-visual physical phenomena and are prone to learning dataset-specific visual biases \cite{ding2024dt}.}

\rev{Audio has emerged as a powerful complementary modality due to its non-destructive, radiation-free, and low-cost nature, and its ability to capture high-frequency mechanical events at a sub-frame temporal resolution. 
Prior work has demonstrated the utility of audio for surgical phase \revThree{and instrument} recognition \cite{seibold2022conditional, fuchtmann_audio-based_2024}, real-time error prevention via breakthrough detection \cite{seibold2021realtime}, or assessing the press-fit of orthopedic implants \cite{goossens2020acoustic}.
However, these methods typically treat audio as a global signal detached from the scene geometry. 
While our previous work introduced a proof-of-concept for 2D sound source localization in image space \cite{seibold2024ssl}, it did not account for the 3D depth and dynamic nature of the surgical environment.}

\rev{
The integration of multimodal data, such as vision, kinematics, language, and audio, is widely recognized as essential for next-generation \ac{cas} \cite{hein2017sds}. 
An early multimodal work by Weede et al. \cite{weede2012workflow} fuses positional information, audio and video signals in a Bayes classifier for surgical phase recognition.
In the wake of recently emerging vision-language models, a series of works fuse vision and language modalities for decision-making in surgical robotics \cite{zargarzadeh2025decision} and workflow analysis \cite{yuan2025learning}.
\revThree{Recent work has also explored audio-visual instrument segmentation, using spoken instrument names to guide visual recognition \cite{chen2024asi}.}
Yet, current approaches generally lack the capability to localize acoustic sources. 
This limitation prevents the system from distinguishing between simultaneous events (e.g., two tools operating at once) or associating a specific sound with a specific instrument in a crowded scene. 
By projecting acoustic localization information from a phased microphone array onto dynamic point clouds, our proposed method bridges this gap, enabling the first spatially aware, multimodal representation of surgical activity.
}

\section{Methodology}

\begin{figure}[t]
    \centering
    \includegraphics[width=\linewidth,height=10cm,keepaspectratio]{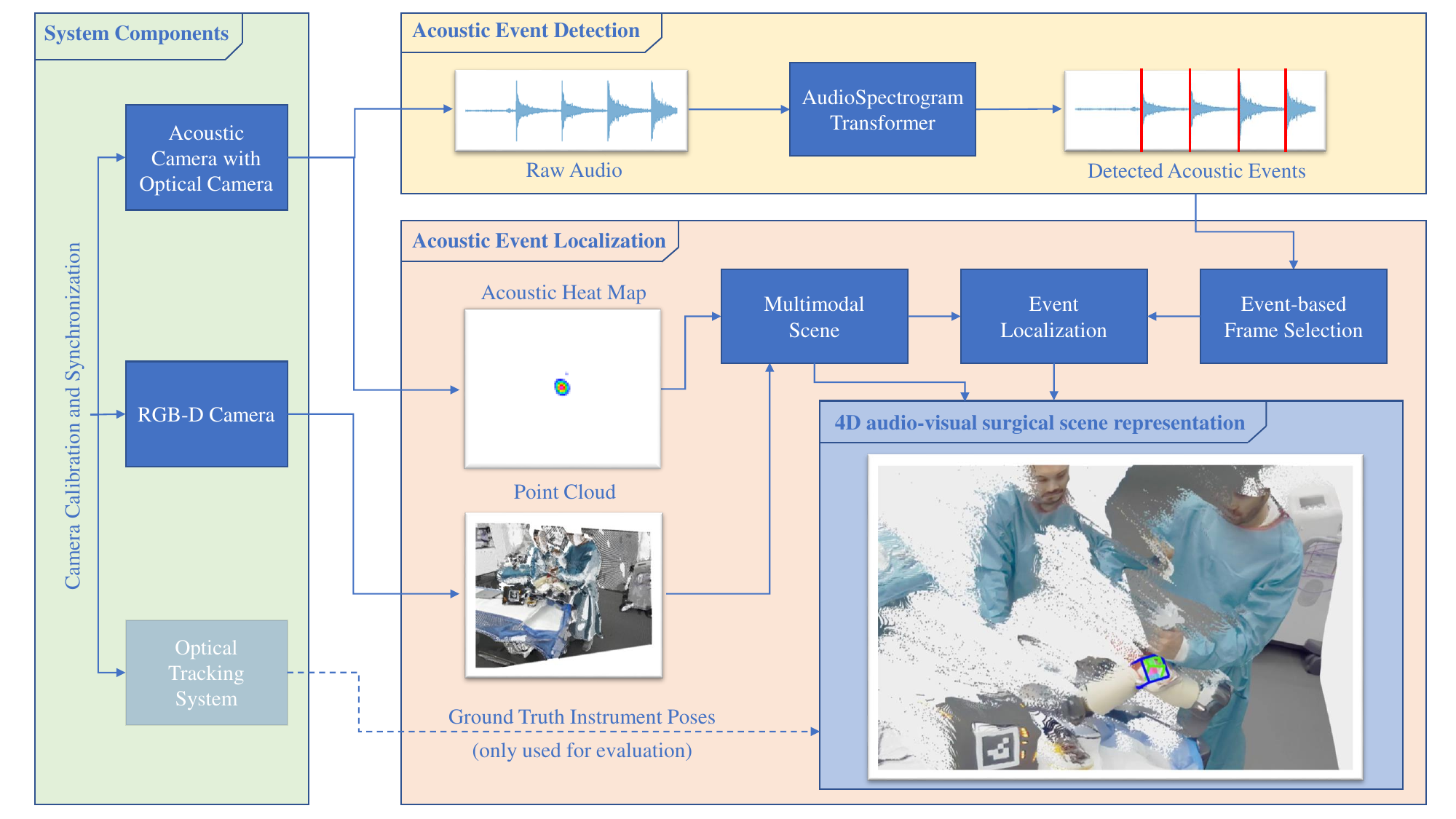}
    \caption{Conceptual overview of the proposed system with an example obtained from surgical chiseling. In this example, hammer events are detected by the \textit{Acoustic Event Detection} stage, as described in \cref{sec:eventdetection}. Detected events trigger the \textit{Acoustic Event Localization} stage, which projects the acoustic heatmap onto the dynamic 3D scene representation and localizes the event. We utilize point clouds from a co-calibrated RGB-D camera to represent the scene in our experiments, as described in \cref{sec:localization}. An optical tracking system is used for the evaluation of the system to compare the result with the ground truth pose of the surgical instrument.}
    \label{fig:overview}
\end{figure}

To create a 4D audio-visual surgical scene representation, our system integrates acoustic information from a phased microphone array, also know as \textit{acoustic camera}, with an existing spatio-temporal scene representation, for example obtained from a RGB-D camera.
As illustrated in \cref{fig:overview}, we propose an acoustic event detection module to detect acoustic events such as the use of surgical instruments, and an event localization module to spatially localize detected events in in the generated multimodal scene representation.
The multimodal scene representation, event detection module, and event localization module are described in \cref{sec:multimodalscene,sec:eventdetection,sec:localization}.

\subsection{Experimental Setup and Data Collection}
\label{sec:experiment}

\begin{figure}
    \hfill
    \includegraphics[width=0.32\linewidth,height=5cm,keepaspectratio]{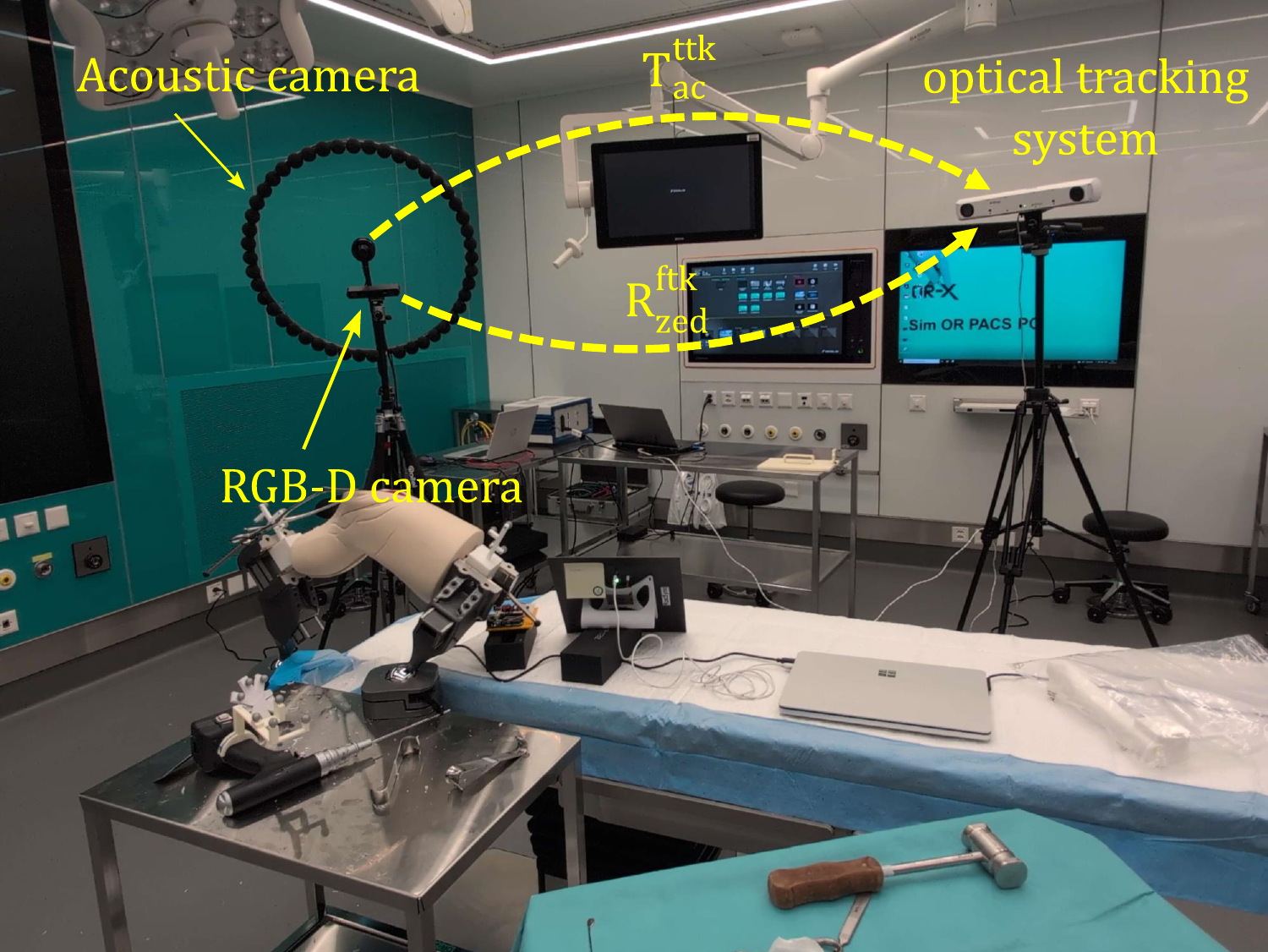}
    \hfill
    \includegraphics[width=0.32\linewidth,height=5cm,keepaspectratio]{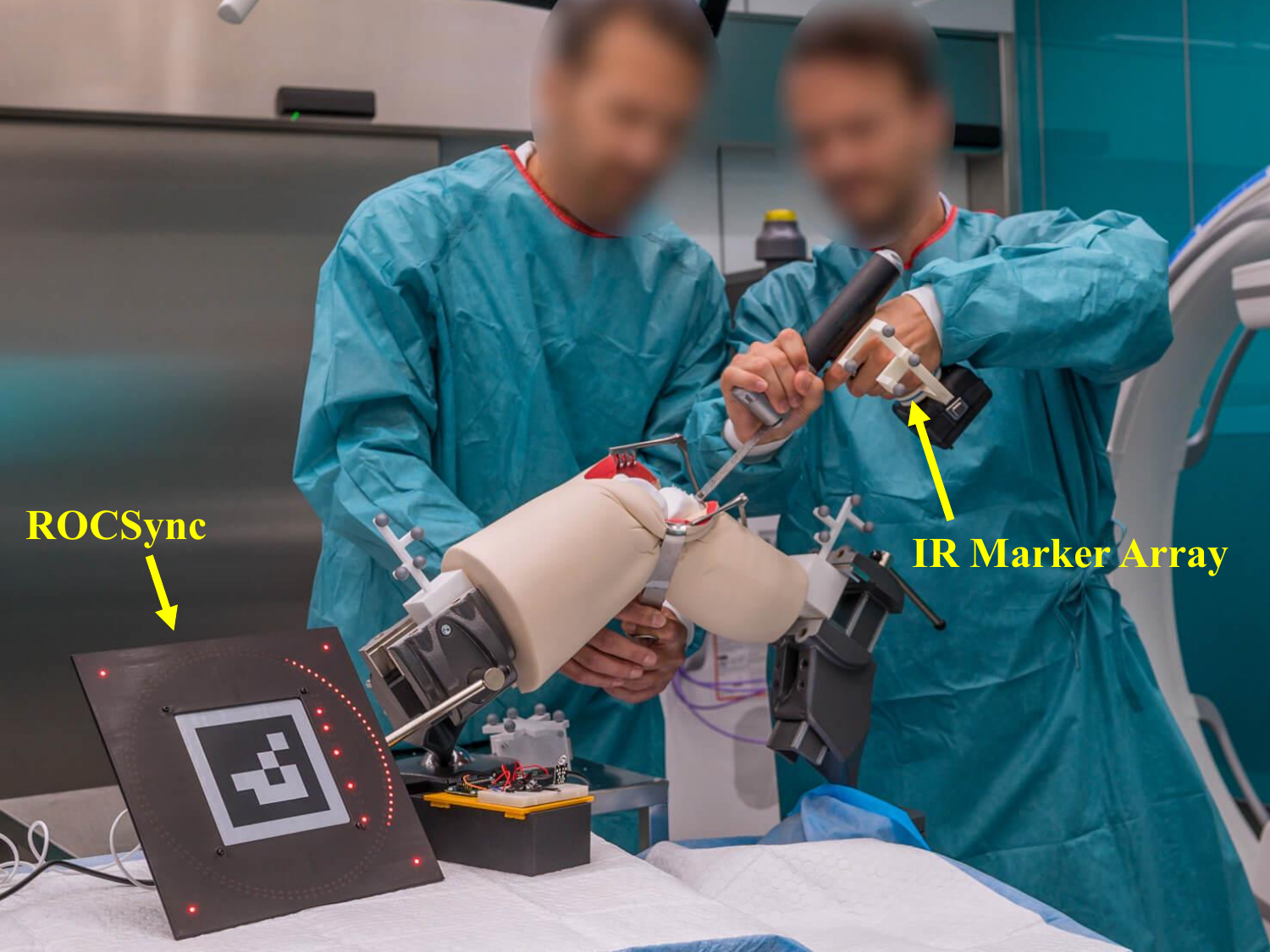}
    \hfill
    \includegraphics[width=0.32\linewidth,height=5cm,keepaspectratio]{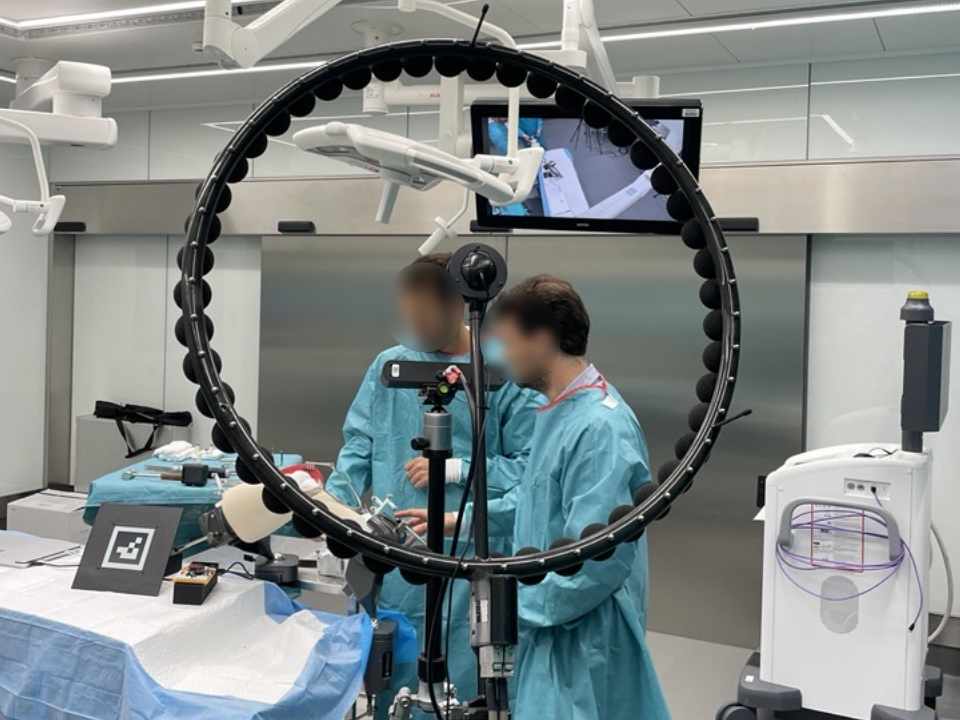}
    \hfill
    \caption{Overview of the experimental setup. We use the Ring48 microphone array, the ZED 2i RGB-D camera, and the FusionTrack 500 tracking system (shown in the left frame) to capture sequences of chiseling, drilling, and sawing (shown in the \revTwo{middle} frame). The right frame shows the viewpoint of the acoustic and RGB-D camera.}
    \label{fig:setup}
\end{figure}

We collected sequences of typical actions that are associated with sound emissions from \ac{tka} using the experimental setup depicted in \cref{fig:setup}.
We simulate surgical sawing, drilling, and chiseling on a plastic bone model including simulated soft tissue \rev{(SYNBONE AG, Zizers, Switzerland)}, conducted by two trained surgical experts with real surgical tools.
For drilling and sawing, we used a Arthrex DrillSaw Max 600 (Arthrex Inc., Naples, USA) surgical-grade power tool. 
We recorded sequences of length $20\:s$ with a frame rate of 25 fps and a sample rate of $192\:kHz$ from the acoustic camera for the surgical actions, resulting in $N_{chiseling} = 6$, $N_{sawing} = 9$, and $N_{drilling} = 5$ sequences with a total length of $400\:s$.

For the presented experiments, we use the gfai tech Ring48 (gfai tech GmbH, Berlin, Germany).
The system employs a phased microphone array with acoustic beamforming to generate two-dimensional heatmaps of acoustic activity. 
An integrated optical sensor allows the acoustic activity map to be superimposed onto an optical image. 
To obtain a dynamic visual 3D representation of the surgical scene, we utilize a state-of-the-art RGB-D camera, namely the ZED 2i (StereoLabs, San Francisco, USA), which captures both color and depth information from the surgical scene. 

To evaluate event localization module, we additionally integrate a FusionTrack 500 optical tracking system (Atracsys, Puidoux, Switzerland), which tracks instrument poses with sub-millimeter accuracy via attached infrared-reflective sphere markers.
The attached marker arrays are calibrated to the system using 3D scans of the tools obtained with a high-fidelity scanner, the Artec Space Spider (Artec3D, Senningerberg, Luxembourg). 
Based on the tracked instrument poses, we compute the ground-truth 3d bounding boxes for event locations.
The ground truth event timestamps for the event detection were labeled manually based on visual and acoustic inspection of the synchronized audio and video recordings. 

\noindent\textbf{System Calibration and Synchronization} All devices are calibrated and synchronized prior to the experiments.
We calibrate the intrinsic parameters of the ZED 2i and Ring48 optical cameras using OpenCV and their extrinsic parameters $T_{ac}^{ftk}$ and $T_{zed}^{ftk}$ relative to the tracking system following the approach of \cite{hein_next-generation_2025}, as shown in \cref{fig:setup}.
To ensure accurate synchronization between all devices throughout the experiments, we utilize a RocSync synchronization device \cite{meyer2026rocsync}, a hybrid clock comprising a linear and binary counter in both the RGB and IR spectrum, which enables a sub-frame accurate synchronization based on timestamps visible in all image frames.
The factory-calibration of the phased microphone array is retained.
At $1920 \times 1080\:px$ resolution, final reprojection errors after calibration and synchronization are $2.30 \pm 2.12\:px$ (Ring48) and $1.52 \pm 0.84\:px$ (ZED 2i).
\label{sec:calibrationandsync}

\subsection{Multimodal Dynamic Surgical Scene Representation}\label{sec:multimodalscene}

We generate a multimodal, spatio-temporal scene representation based on the fusion of the acoustic information with the dynamic 3D point cloud.
We compute 2D acoustic heatmaps with a fixed size of $100 \times 100\:px$ via time-domain beamforming, using the professional sound analysis software NoiseImage 4.16.0 (gfai tech GmbH, Berlin, Germany). 
The heatmaps are computed at a predefined distance from the phased microphone array using a fixed 2D scanning grid. 
Band-filtering is applied between $1$ and $5\:kHz$ for sawing and between $1$ and $10\:kHz$ for drilling to obtain clear heatmaps without artifacts from ambient sound. 
For chiseling, no filtering is applied. 
The acoustic heatmaps are fused with the 3D scene representation by a simple projection operation.
We project the amplitude of the acoustic heatmap, normalized to the range $[0, 1]$.

\subsection{Acoustic Event Detection}\label{sec:eventdetection}

Our proposed acoustic event detection stage operates directly on the raw audio captured by the acoustic camera. 
We resample the audio with a sample rate of $SR=16\:kHz$ samples per second, apply a sliding window with length $L_W = 150\:ms$ and hop length $L_{hop} = 20\:ms$ to the sequence, and compute mel spectrograms with $n_{mels} = 128$ mel bins from each window. 
\rev{For event detection, we fine-tune a pretrained AudioSpectrogramTransformer (AST) \cite{gong21b_interspeech} in a multi-class classification task (classes = {idle, chiseling, drilling, sawing}) to predict event presence using PyTorch 2.8.0.} 
During training we apply audio data augmentations, i.e., gain variation, gaussian noise, gain transitioning, clipping distortion, time stretching, and pitch shifting. 
The outputs are converted to video timestamps by detecting transitions in model predictions, as described in detail in the evaluation in \cref{sec:eventdetectionresults}. 
For sawing and drilling, we export multiple acoustic image frames in regular intervals after a detected event to evaluate the localization stage, as described in \cref{sec:localizationresults}.

\subsection{Event Localization}\label{sec:localization}

The location of a detected sound source is approximated within a 3D bounding box based on the computed multimodal scene representation.
We localize areas with high signal amplitude within the scene using weighted clustering.
Specifically, we use DBSCAN \cite{ester1996density} with a neighborhood radius $r = 30.0\:mm$, a minimum cluster weight of $\rev{w = 400.0}$, and the acoustic signal amplitudes as per-point weights.
A tight bounding box is computed around the cluster with the largest total weight.
Last, we limit the \rev{side lengths of the predicted bounding box} to a predefined range, based on the instrument size and the acoustic characteristics of each action.
For drilling and sawing, the sound originates from the power tool surface due to motor-induced vibrations. 
\rev{Therefore, we limit the box side lengths to the minimum and maximum side length of the instrument's ground-truth bounding box.}
In contrast, for chiseling, sound emission is concentrated at the contact edge resulting from collisions between the mallet/chisel and the chisel/bone anatomy. 
Thus, we define a fixed-size bounding box with an edge length of 5 cm centered on the chisel's contact edge.
\rev{Since this step is executed after the event detection, the approximate physical dimensions of the sound-emitting instruments can be assumed to be known.}

\section{Evaluation}\label{sec:evaluation}

\begin{figure}[t]
\centering

\begin{subfigure}{.33\textwidth}
    \centering
    \includegraphics[width=\linewidth]{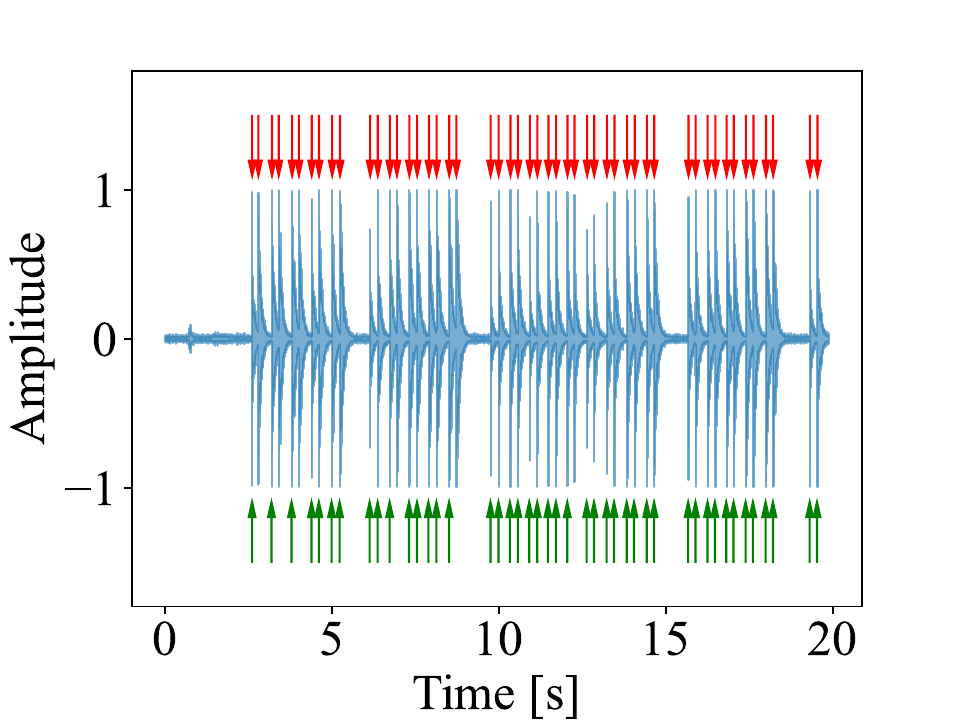}
    \caption{Chiseling}
\end{subfigure}%
\begin{subfigure}{.33\textwidth}
    \centering
    \includegraphics[width=\linewidth]{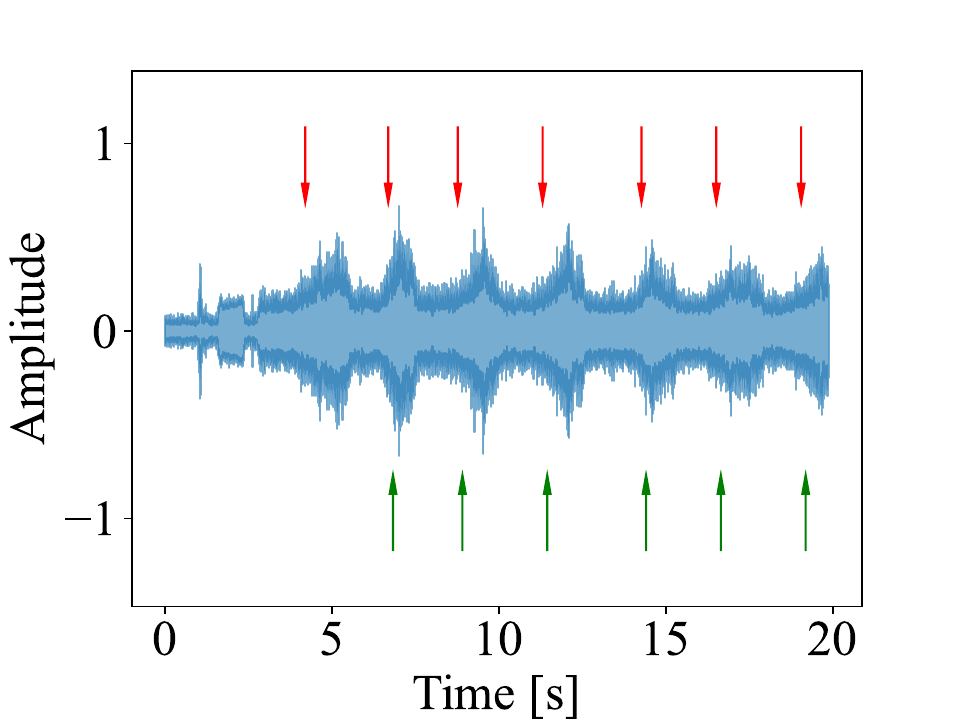}
    \caption{Drilling}
\end{subfigure}%
\begin{subfigure}{.33\textwidth}
    \centering
    \includegraphics[width=\linewidth]{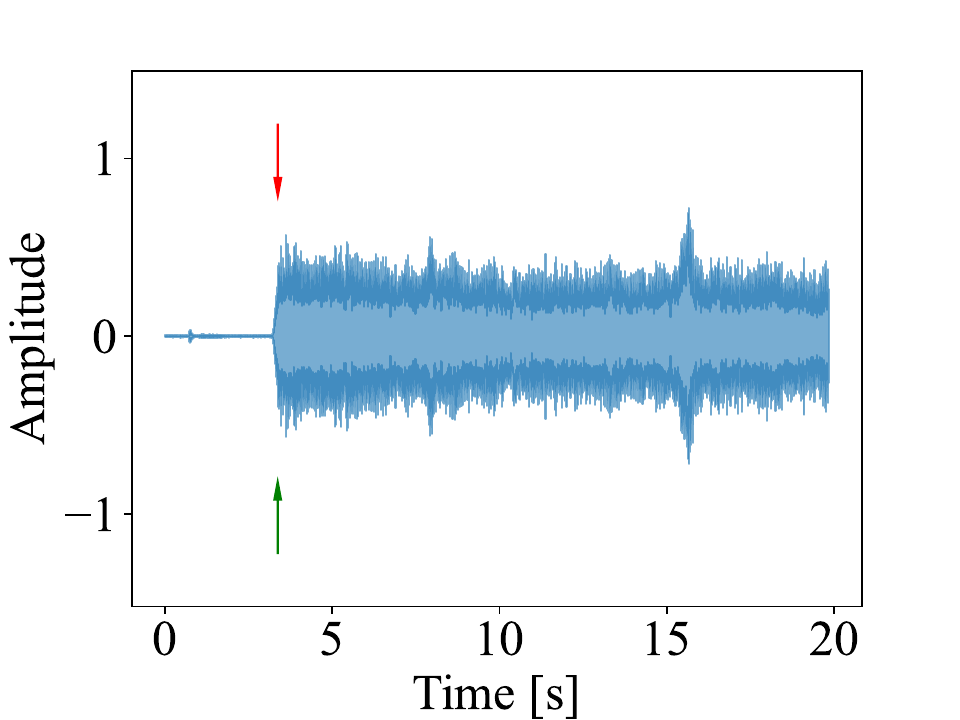}
    \caption{Sawing}
\end{subfigure}

\caption{We illustrate qualitative results over one full example recording of length $L_{sequence} = 20s$, obtained from the experiment described in \cref{sec:evaluation}, for each of the surgical actions analyzed within this work. The detections have been obtained using the relaxed condition. The red arrows on top are the ground truth events, the green arrows on the bottom are the predicted events.}
\label{fig:eventdetectionresults}
\end{figure}

\noindent\textbf{Acoustic Event Detection} \cref{tab:metrics} illustrates the quantitative results for the proposed event detection method described in \cref{sec:eventdetection}. All results have been obtained using \rev{3 -fold cross validation}. We detect an event as a change in sign from $y_{pred} = 0$ to $y_{pred} = 1$ in the model predictions, where $y_{pred} = 0$ corresponds to no event detected and $y_{pred} = 1$ corresponding to a detected event. We introduce a \textit{relaxed} condition, that allows for an event to be detected within a $20\:ms$ window of $j$ frames around the ground truth event to be counted as a true positive prediction. We set this parameter as $j_{chiseling} = 1$, \rev{$j_{sawing} = j_{drilling} = 10$}. The \textit{hard} condition only accepts predictions in the correct frame according to the ground truth. \cref{fig:eventdetectionresults} shows qualitative results for full example recordings.
\label{sec:eventdetectionresults}

\begin{table}[t]
\centering
\caption{\rev{Precision, Recall, and F1 under hard and relaxed conditions}}
\begin{tabularx}{\textwidth}{l|*{2}{>{\centering\arraybackslash}X}|*{2}{>{\centering\arraybackslash}X}|*{2}{>{\centering\arraybackslash}X}}
\toprule
\textbf{Method} & 
\multicolumn{2}{c|}{\textbf{Precision [\%]}} & 
\multicolumn{2}{c|}{\textbf{Recall [\%]}} & 
\multicolumn{2}{c}{\textbf{F1 [\%]}} \\
\cmidrule(lr){2-3} \cmidrule(lr){4-5} \cmidrule(lr){6-7}
 & \textbf{Hard} & \textbf{Relaxed} & \textbf{Hard} & \textbf{Relaxed} & \textbf{Hard} & \textbf{Relaxed} \\
\midrule
Chiseling & $0.915 \pm 0.045$ & $\mathbf{0.968} \pm \mathbf{0.022}$ & $0.791 \pm 0.134$ & $\mathbf{0.953} \pm \mathbf{0.033}$ & $0.845 \pm 0.093$ & $\mathbf{0.961} \pm \mathbf{0.026}$ \\
Drilling  & $0.000 \pm 0.000$ & $\mathbf{0.753} \pm \mathbf{0.194}$ & $0.000 \pm 0.000$ & $\mathbf{0.639} \pm \mathbf{0.160}$ & $0.000 \pm 0.000$ & $\mathbf{0.651} \pm \mathbf{0.077}$ \\
Sawing    & $0.611 \pm 0.283$ & $\mathbf{1.000} \pm \mathbf{0.000}$ & $0.390 \pm 0.079$ & $\mathbf{0.889} \pm \mathbf{0.157}$ & $0.444 \pm 0.079$ & $\mathbf{0.933} \pm \mathbf{0.094}$ \\
\bottomrule
\end{tabularx}
\label{tab:metrics}
\end{table}

\begin{figure}[t]
    \centering
    \includegraphics[width=.49\linewidth,height=4cm,keepaspectratio]{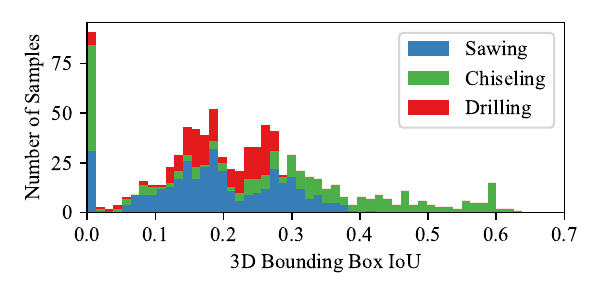}%
    \includegraphics[width=.49\linewidth,height=4cm,keepaspectratio]{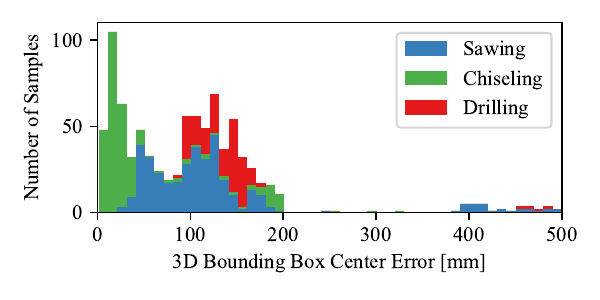}
    \caption{Distribution of 3D bounding box IoU scores \rev{and 3D bounding box center errors} over a total of 20 recordings, grouped by surgical action.}
    \label{fig:localizationhistogram}
\end{figure}

\noindent\textbf{Acoustic Event Localization} To evaluate our event localization approach, we compute the 3D \ac{iou} of the predicted and ground-truth 3D bounding boxes. 
For drilling and sawing actions, the ground-truth bounding box is computed from the instrument mesh and tracked pose obtained from the marker-based tracking system.
For chiseling sequences, \revThree{the ground-truth is defined as a} bounding box with a side length of $5\:cm$ centered on the chisel's contact edge, matching the definition used during inference.
\rev{The distribution of IoU scores is shown in \cref{fig:localizationhistogram} where a IoU threshold of 0.1 results in recall rates of 0.78, 0.91, 0.84, and 0.84 for chiseling, drilling, sawing, and \revTwo{overall} average, respectively. 
For further insights,} please refer to the supplementary material, where we evaluate the localization recall rates under multiple of \ac{iou} thresholds ranging from $0.05$ to $0.40$.
Further qualitative results are provided in the supplementary videos.

\label{sec:localizationresults}

\subsection{\rev{Ablation Studies}}
\begin{figure}
    \centering
    \includegraphics[width=.6\linewidth]{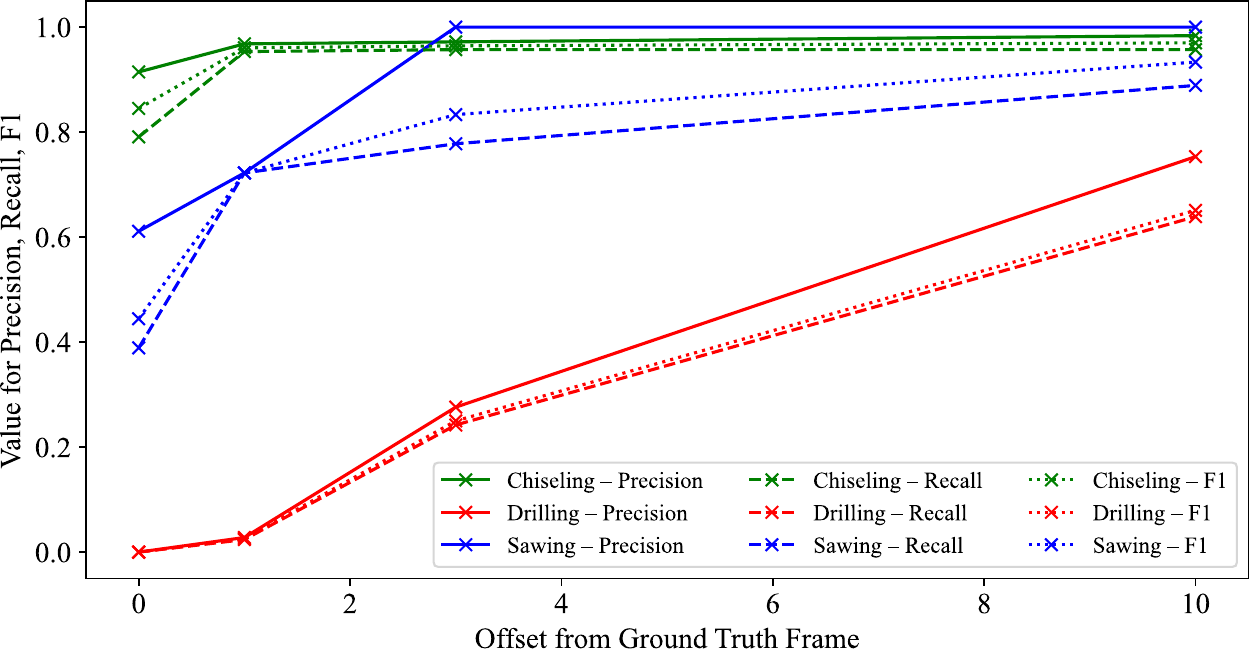}
    \caption{\rev{The onset error distribution for the event detection component.}}
    \label{fig:onseterror}
\end{figure}

\rev{To provide further insights into the result distribution, we perform a detailed analysis of the onset error distribution for the event detection stage of the proposed method and present the results in figure \ref{fig:onseterror}. Furthermore, we evaluate the robustness of the event detection stage under simulated realistic OR background noise in the supplementary material.
In \cref{fig:dbscan_sensitivity_analysis}, we analyze the sensitivity in the choice of DBSCAN parameters, as well as the density of the input point cloud. 
We vary one parameter at a time in the ranges of $r\in [7.5, 60]\:mm$ and $w\in [50, 800]$.
In addition, we analyze the robustness to the input point cloud's density by downsampling it by a factor of $s \in [\frac{1}{64}, 1]$.
}

\rev{Last, we compare our localization stage against a naive centroid detection baseline, which computes the 3D sound source location as the weighted average of the entire point cloud, using the sound amplitude as per-point weights. To obtain a region-of-interest bounding box, an axis-aligned bounding box of the instrument's size is centered around the estimated location.}
\rev{This baseline achieves a 3D bounding box IoU of $0.14 \pm 0.16$ compared to $0.23 \pm 0.14$ of our localization stage. 
The 3D bounding box center error of this baseline is $144.10 \pm 121.19\:mm$ compared to $101.39 \pm 89.75\:mm$ obtained by the proposed localization stage.}

\begin{figure}[t]
\centering
\begin{subfigure}{.33\textwidth}
    \centering
    \includegraphics[width=\linewidth]{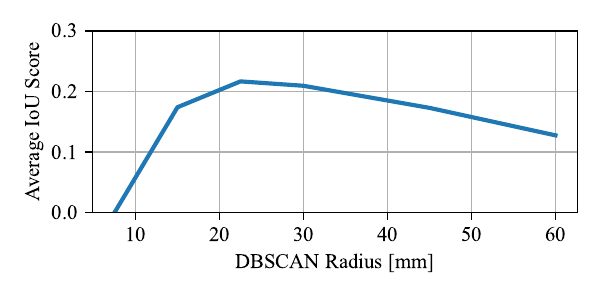}
\end{subfigure}%
\begin{subfigure}{.33\textwidth}
    \centering
    \includegraphics[width=\linewidth]{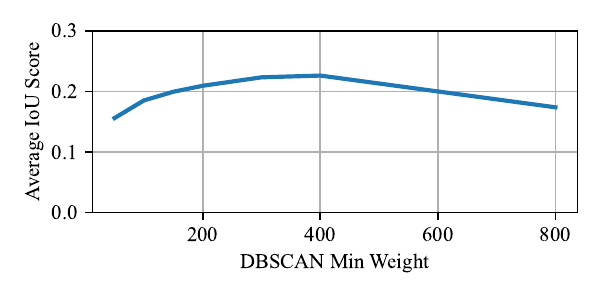}
\end{subfigure}%
\begin{subfigure}{.33\textwidth}
    \centering
    \includegraphics[width=\linewidth]{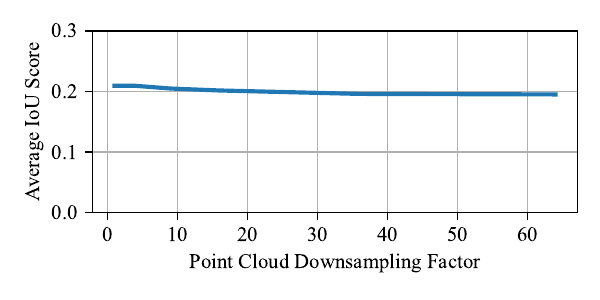}
\end{subfigure}
\caption{\rev{Sensitivity analysis of the DBSCAN radius and minimum weight parameters. We additionally evaluate the effect of the point cloud density on the localization accuracy by downsampling the input point cloud, while adjusting the minimum weight accordingly. All results are reported as the average IoU on the whole dataset.}}
\label{fig:dbscan_sensitivity_analysis}
\end{figure}

\section{Discussion}

In this paper, we introduce a novel concept that enables the generation of spatio-temporal and audio-visual surgical scene representations by mapping acoustic features generated by surgical actions to a dynamic point cloud of the scene. 
\revThree{
In this context, the presented work is the first to map acoustic information into digital surgical scene descriptions. 
This audio-visual representation can directly benefit downstream tasks such as action recognition, phase recognition, and workflow analysis. 
Vision-based approaches typically rely on instrument presence as a discriminative feature \cite{czempiel_opera_2021}, but cannot distinguish tool proximity from tool-tissue interaction.
The acoustic signal encodes these instrument states, providing complementary information that is largely inaccessible to visual analysis alone. 
In the context of surgical scene graphs, localized and classified sound sources enable the modeling of richer semantic relations, for example by augmenting spatial entity relations with interaction-level predicates derived from acoustic activity.
Additionally, acoustic features provide complementary information to disambiguate visually similar instruments, such as the drill and saw attachments in our experiments, which share the same power tool basis. 
}

The event detection module showed promising results, especially for chiseling and sawing, where events could be detected within a maximum offset of one frame for chiseling (corresponding to at most $20\:ms$). 
This temporal resolution is particularly important for sharp burst events such as chiseling. 
For drilling and sawing, the actions span over a longer period of time, as visible in the waveform plot in \cref{fig:eventdetectionresults}, making a relaxed onset criterion tolerable.
The detection of drilling events presented a more challenging problem which can mainly be attributed to the fact that the surgeons turned on the device in the beginning of the sequence and kept drilling into the plastic bone while the power tool was running, resulting in only very subtle sound differences between drilling in the air and drilling into the rather soft plastic bone. To mitigate this issue, beam steering or noise canceling could improve the results, which is, however, beyond the scope of the presented work.

The event localization module proved that the fusion of audio-visual information is an effective approach to detect regions of interest with minimal computational overhead.
Despite the simplicity of our approach, $\revTwo{84 \%}$ of the events were successfully localized with a 3D bounding box \ac{iou} of at least $0.1$. 
\revThree{
While this level of accuracy is sufficient for coarse downstream tasks, more precise tasks such as fine-grained instrument tracking would demand substantially tighter localization. 
The steep decrease in recall under stricter \ac{iou} thresholds reflects both the low resolution of the acoustic camera and the constraints of the learning-free localization baseline.
}

\rev{Our implementation has a latency of about $250\:ms$ for the event detection and about $100\:ms$ for the localization stage (measured on a downsampled point cloud with $\frac{1}{64}$ of the original size). 
While we cannot measure the latency of the beam-forming algorithm due the absence of an API in the closed-source software, comparable methods take around $200\:ms$ \cite{jansen2026delay}. 
Our current implementation only supports offline processing for the same reason.
}

The exploratory character of our experiments inherently comes with some limitations. 
\rev{The presented dataset is limited in scale \revTwo{and diversity}, containing sequences of three representative surgical action, and does not represent the full complexity of a real surgery. 
Still, our preliminary results show the potential of our novel multimodal audio-visual surgical scene representation and motivate further research to investigate various aspects for improvement, e.g., a wider range of surgical activity, advanced beamforming algorithms, high-quality 3D scene representations and their enrichment via the integration of additional low-level perception algorithms. 
During the experiment, we used a plastic bone and foam soft tissue model that was moved frequently to introduce variation. 
The robustness of the proposed learning-free localization approach should be subject to future studies that should evaluate realistic patient models, as well as obtaining larger amounts of data with more variation.}
Another limitation is the use of a single-view point cloud from a ZED2i RGB-D camera as the dynamic 3D visual representation.
While this produces reasonable quality point clouds, the preservation of fine details and scene completeness is inherently limited by single-view capture and associated occlusions.

More broadly, acoustic information has been shown to have great potential for the analysis of surgical activity in related work, however, acoustic signal analysis is inherently limited to actions that generate audible sounds.
While this makes the approach highly effective in sound-rich fields like orthopedics, its application to other surgical sub-fields may be more challenging.
However, the proof-of-concept presented in this work does not fully utilize the high sensitivity of the acoustic camera. 
Especially for hard-to-detect sounds, acoustic beam steering can be a highly promising solution to improve the sound quality for acoustics-based downstream tasks. 

\revThree{
Several directions for future research emerge from this work.
First, improving localization accuracy is critical for enabling fine-grained downstream tasks. 
This could be achieved by replacing the learning-free baseline with a deep learning-based approach that extracts more descriptive multimodal features, and by evaluating advanced beamforming algorithms beyond the standard delay-and-sum method used in this study.
Second, the visual representation should be upgraded through multi-view reconstructions to reduce occlusions, and through alternative representations such as dynamic Gaussian splats, which could provide higher-fidelity scene geometry.
Third, the experimental scope should be expanded to include a wider range of surgical actions, realistic patient models, and larger datasets with greater variation, to validate the robustness and generalizability of the proposed approach.
Finally, further optimizations such as GPU-based acceleration, an open beamforming pipeline, and online processing would be necessary steps toward real-time intraoperative deployment.
}

\section{Conclusion}

In summary, this work introduces a novel framework for surgical scene understanding that fuses visual and acoustic information into a unified 4D audio-visual digital representation. By combining RGB-D–based dynamic point clouds with sound localization information obtained from a phased microphone array, and by incorporating transformer-based acoustic event detection, our approach not only enhances contextual awareness in surgical environments but also lays the groundwork for future intelligent and autonomous systems capable of richer, data-driven surgical scene interpretation.

\section*{Declarations}

\noindent\textbf{Acknowledgements}
This work has been funded by the Swiss National Foundation under the funding scheme "Spark" with grant number 228813. 
The experiments were supported by the OR-X - a Swiss national research infrastructure for translational surgery and associated funding by the University Hospital Balgrist. 
Imaging was performed with support of the Swiss Center for Musculoskeletal Imaging (SCMI). 
We thank Dr. Matthias Brechbühl from Norsonic Brechbühl AG and Carsten Hessenius, Andy Meyer, and Michael Markus Ackermann from GFaI e.V. and gfai tech GmbH for their generous support.

\noindent \textbf{Competing Interests} The authors have no relevant financial or non-financial interests to disclose.

\noindent \textbf{Ethics Approval}
Since this study did not involve human or animal subjects, ethics approval is not applicable.

\noindent \textbf{Informed Consent}
As this research did not involve human participants, informed consent is not applicable.

\bibliography{bibliography}

\end{document}